\documentclass[prb,altaffillsymbol,superscriptaddress, 
citeautoscript,amsfonts,amssymb,amsmath,twocolumn,floatfix,altaffillsymbol]
{revtex4}

\usepackage{graphicx}
\begin{document}

\title{Metallic conductivity and a Ca substitution study of NaRh$_2$O$_4$ comprising a double chain system}

\author{K. Yamaura}
\email[E-mail at:]{YAMAURA.Kazunari@nims.go.jp}
\homepage[Fax.:]{+81-29-860-4674}
\affiliation{Superconducting Materials Center, National Institute for Materials Science, 1-1 Namiki, Tsukuba, Ibaraki 305-0044, Japan}

\author{Q. Huang}
\affiliation{NIST Center for Neutron Research, National Institute of Standards and Technology, Gaithersburg, Maryland 20899, USA}

\author{D.P. Young}
\affiliation{Department of Physics and Astronomy, Louisiana State University,
Baton Rouge, LA 70803}

\author{M. Moldovan}
\affiliation{Department of Physics and Astronomy, Louisiana State University,
Baton Rouge, LA 70803}

\author{X.L. Wang}
\affiliation{Institute for Superconducting and Electronic Materials, University of Wollongong, Wollongong, NSW 2522, Australia}  

\author{E. Takayama-Muromachi}
\affiliation{Superconducting Materials Center, National Institute for Materials
Science, 1-1 Namiki, Tsukuba, Ibaraki 305-0044, Japan}

\begin{abstract}

The metallic compound $\rm NaRh_2O_4$ forms a full range solid solution to the insulating phase $\rm CaRh_2O_4$.  At a Na concentration of 0.25 moles per formula unit, we found an unexpected contribution to the specific heat at low temperature [K. Yamaura et al. Chem. Mater. 17 (2005) 359].  To address this issue, specific heat and ac and dc magnetic susceptibilities were additionally measured under a variety of conditions for the $\rm Na_{0.25}$ sample.  A new set of data clearly indicate the additional specific heat is magnetic in origin; however, the magnetic entropy is fairly small ($\sim$1 $\%$ of Schottky term for a simple splitting doublet), and there is no other evidence to suggest that a magnetic phase transition is responsible for the anomalous specific heat.

\end{abstract}

\maketitle

\begin{figure}[!ht]
\begin{center}
\includegraphics[width=0.40\textwidth]{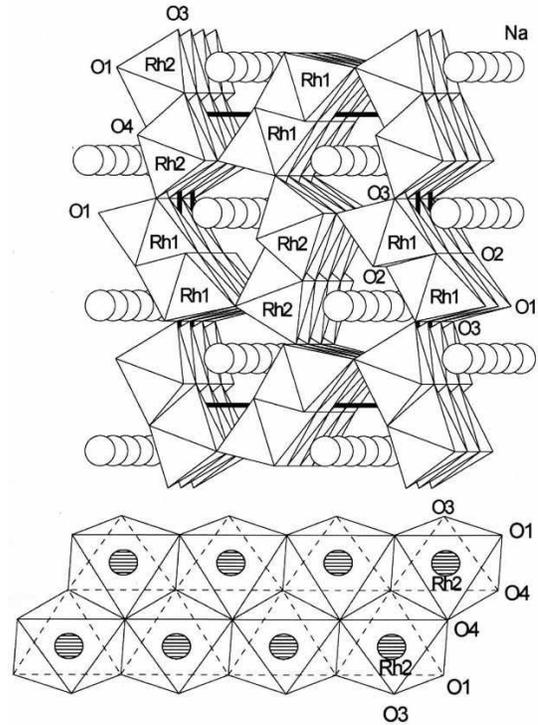}
\end{center}
\caption{Structure view of $\rm NaRh_2O_4$.  The lower figure clarifies the double chain detail along the $b$ axis. }
\label{fig1}
\end{figure}

The metallic oxide $\rm NaRh_2O_4$ was recently synthesized for the first time by a high-pressure method in Tsukuba, Japan \cite{1}.  The oxide crystallizes in the $\rm CaFe_2O_4$-type structure (Fig.\ref{fig1}) which comprises a double chain system.  All Rh ions are in the octahedral environment (oxygen coordinated) and the octahedra RhO$_6$ constitute the chain system by sharing those edges.  It is by no means an infrequent occurrence to see such an octahedra network which is not the perovskite type and shows fairly metallic behavior.  $\rm NaRh_2O_4$ shows also a broad cusp in its static magnetic susceptibility vs. temperature curve at approximately 100 K, implying many possibilities of interesting magnetism.  

In this work, the aliovalent substitution Na$^+$/Ca$^{\rm 2+}$ was studied.  Solid-solution compounds in the whole Na/Ca range were successfully prepared in the same chemical manner, and the unit-cell volume was found to vary monotonically with the substitution as well as with the formal valence of the Rh ion, reduced from +3.5 (Na) to +3.0 (Ca).  In detail, no structure transformation was detected in the chemical range, and the change in the unit cell volume was small, 1.7 $\%$.  As the trivalent Rh ion in the octahedral environment is not magnetically active ($4d^6: t_{\rm 2g}^6 e_{\rm g}^0$), the chemical substitution should alter the magnetic character.  

Results thus far have revealed an unexpected term in the specific heat at a Na concentration of 0.25 per formula unit \cite{1}.  Here, we focused on the 0.25-Na sample and conducted measurements of specific heat in a magnetic field of 70 kOe, as well as ac and dc susceptibilities in a weak magnetic field.  

The same sample previously prepared at $\rm Na_{0.25}$ was reemployed for additional studies \cite{1}.  Specific heat of a small piece of the pellet ($\sim$30 mg) was measured in a commercial apparatus (Q.D. PPMS) between 1.8 and 20 K in a field of 70 kOe.  DC magnetic susceptibility was measured in a commercial apparatus (Q.D. MPMS) between 1.8 K and 150 K at 10 Oe on heating after cooling without the field and on cooling with the field.  The residual magnetic field in the sample space was carefully nullified before the measurements.  AC magnetic susceptibility was measured on heating from 5 K to 300 K at 1 Oe (frequencies: 21, 117, 217, 500, 2000, 5000, 8000 Hz) in a commercial apparatus (Q.D. PPMS).

\begin{figure}[!ht]
\begin{center}
\includegraphics[width=0.40\textwidth]{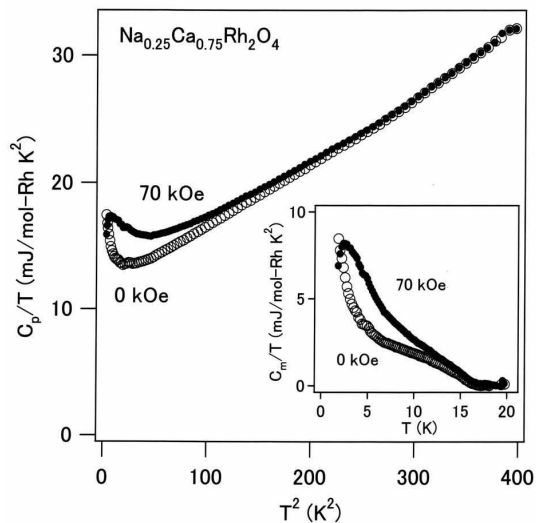}
\end{center}
\caption{Specific heat data for $\rm Na_{0.25}Ca_{0.75}Rh_2O_4$ measured in a magnetic field of 70 kOe and at zero field.  (Inset) Magnetic contribution to each of the data.}
\label{fig2}
\end{figure}

Fig.\ref{fig2} shows the specific-heat data for $\rm Na_{0.25}Ca_{0.75}Rh_2O_4$ plotted as $C_{\rm p}/T$ vs. $T^2$.  A small upturn is clearly seen at zero field.  The feature was only observed for the Na$_{\rm 0.25}$ sample [1].  As we speculated the upturn might be magnetic in origin, the magnetic field dependence was studied to examine the possibility.  After the 70 kOe field was applied, the upturn was found to shift to the right, and a sharp maximum appeared beyond the experimental limit.  This qualitative behavior does suggest the upturn is related to the sample magnetism. 

For the 70 kOe data, we tried to roughly estimate the magnetic entropy, although some of the area below the peak was outside of the measurement range.  A simple formula was applied to the data in the previous study \cite{1}, and electronic and lattice contributions were carefully subtracted from the data.  The resultant data were plotted against temperature in the inset of Fig.\ref{fig2}.  We found the magnetic entropy to be approximately 50 mJ/K mole of Rh.  This value is fairly small when compared to the entropy value for what is expected for a Schottky term (splitting of a simple doublet).  The result amounts to approximately 1$\%$ of Rln2 (R: universal gas constant).  Although qualitatively the behavior of the anomalous contribution to the specific heat appears magnetic in origin, quantitatively speaking, the very small value of the entropy makes it difficult to verify this hypothesis, since the term is too small for a fitting study using various magnetic models.  Probably, diluted paramagnetic spins, caused by the chemical substitution, are involved in the magnetic origin, which might be complicated somewhat by a relatively weak force of localization of the itinerant electrons \cite{1}.  

In summary, a specific heat study of the $\rm Na_{0.25}$ sample was conducted in magnetic field, and the additional term was suggested to be a magnetic one.  Its entropy was, however, too small to be addressed properly by a fitting study.  To further test the possibility of a magnetic transition being responsible for the magnetic term, magnetic susceptibilities were carefully measured in weak ac and dc fields (data are not shown here), and the results indicated this possibility was unlikely. Further studies would be required to make clear the origin of the magnetic term at $\rm Na_{0.25}$ and to establish and understand the whole picture of the magnetic character of $\rm NaRh_2O_4$. 

This research was supported in part by the Superconducting Materials Research Project, administrated by MEXT, and by Grants-in-Aid for Scientific Research from JSPS (16076209, 16340111).


\begin{thebibliography}{99}

\bibitem{1} K. Yamaura et al. Chem. Mater. {\bf 17} (2005) 359.  

\end{thebibliography}
\end{document}